\documentclass[twocolumn,prl,floatfix,citeautoscript,nofootinbib]{revtex4}
\usepackage{amsbsy}
\usepackage{latexsym,epsfig,graphicx}
\usepackage{dcolumn}
\usepackage{graphicx}
\usepackage{subfigure}
\usepackage{comment}
\usepackage{color}
\usepackage{bm}
\usepackage{mathrsfs}
\usepackage{amssymb}
\usepackage[colorlinks,urlcolor=blue,citecolor=blue]{hyperref}

\begin{document}

\author{Junpeng Hou}
\author{Haiping Hu}
\author{Kuei Sun}
\author{Chuanwei Zhang}
\thanks{Corresponding author. \\
Email: \href{mailto:chuanwei.zhang@utdallas.edu}{chuanwei.zhang@utdallas.edu}%
}
\affiliation{Department of Physics, The University of Texas at Dallas, Richardson, Texas
75080-3021, USA}
\title{Superfluid-quasicrystal in a Bose-Einstein condensate }

\begin{abstract}
Quasicrystal is a class of ordered structures defying conventional
classification of solid crystals and may carry classically forbidden (e.g.,
5-fold) rotational symmetries. In view of long-sought supersolids, a natural
question is whether a superfluid can spontaneously form quasicrystalline
order that is not possessed by the underlying Hamiltonian, forming
``superfluid-quasicrystals". Here we show that a superfluid-quasicrystal
stripe state with the minimal 5-fold rotational symmetry can be realized as
the ground state of a Bose-Einstein condensate within a practical
experimental scheme. There exists a rich phase diagram consisting of various
superfluid-quasicrystal, supersolid, and plane-wave phases. Our scheme can
be generalized for generating other higher-order (e.g., 7-fold) quasicrystal
states, and provides a platform for investigating such new exotic quantum
matter.
\end{abstract}

\maketitle

\emph{Introduction}. Quasicrystals exhibit exotic spatial patterns that are
neither periodic as solid crystals (\textit{i.e.}, lack of translational
symmetry) nor totally disordered (\textit{i.e.}, possession of long-range
order) \cite{Christian1994}. The Bragg diffraction peaks of quasicrystals
possess rotational symmetries such as $5$, $7$, $8$, $9$, $10$-fold that are
forbidden in classical crystalline orders \cite{Christian1994, Levine1984}.
Since its first report in Al-Mn and Al-Mn-Si alloys in 1984 \cite%
{Shechtman1984}, quasicrystal order has been studied and discovered in many
different materials \cite{Macia2006, Freedman2007, Barkan2011, Wasio2014,
Nagao2015, Urgel2016, Bindi2009}.

Supersolid, another exotic phase of matter, combines solid crystalline
structure with superfluidity, where two continuous symmetries, namely,
translational and $U(1)$ gauge, are spontaneously broken \cite%
{Boninsegni2012}. Supersolids were first predicted for helium almost $50$
years ago \cite{Thouless1969, Andreev1971}, and have recently been observed
in cold atom experiments \cite{Li2017, Leonard2017}, where a stripe phase
with supersolid properties was generated and observed in a Bose-Einstein
condensate (BEC) \cite{Li2017}. These great advances in the study of
supersolids raise a natural question: is it possible to create a novel
quantum matter where both superfluidity and quasicrystal orders coexist?

In this Letter, we address this important question by proposing a scheme to
generate a stable quasicrystal ground state in a BEC. The experimental setup
contains a 3D BEC confined in a 1D optical superlattice with quintuple wells
(defines 5 pseudospin states), where neighboring wells are coupled by Raman
assisted tunneling to generate an effective spin-orbit coupling (SOC) \cite%
{Li2016, Li2017} in the perpendicular plane. The scheme utilizes natural
contact interaction and can realize quasicrystals with the minimum 5-fold
rotational symmetry. In this new quantum state, the $U(1)$ gauge symmetry is
spontaneously broken just as that in supersolid stripe phases \cite{Li2016,
Li2017}. However, the discrete translational symmetry, which is preserved in
supersolids and leads to periodic density modulations in stripe phases \cite%
{Li2016, Li2017}, has also been broken, leaving only specified rotational
symmetry. A quasicrystal order with such rotational symmetry but no periodic
spatial density modulation is spontaneously formed although the underlying
Hamiltonian does not possess such order. Therefore we denote this quantum
matter as `superfluid-quasicrystal'. By tuning system parameters (e.g.,
Raman coupling strength, detuning, interaction, etc.), we show, through both
variational ansatz analysis and direct simulation of mean field
Gross-Pitaevskii equation (GPE), that there exists a rich phase diagram
containing various superfluid-quasicrystals, supersolids, and plane-wave
phases. Our scheme can be further extended to generate any $n$-order
superfluid-quasicrystal phases. Our results may advance our understanding of
both quasicrystals and superfluids and should provide an excellent platform
for exploring many interesting properties of superfluid-quasicrystals, a
novel format of quantum matter.

\begin{figure}[t]
\centering
\includegraphics[width=0.45\textwidth]{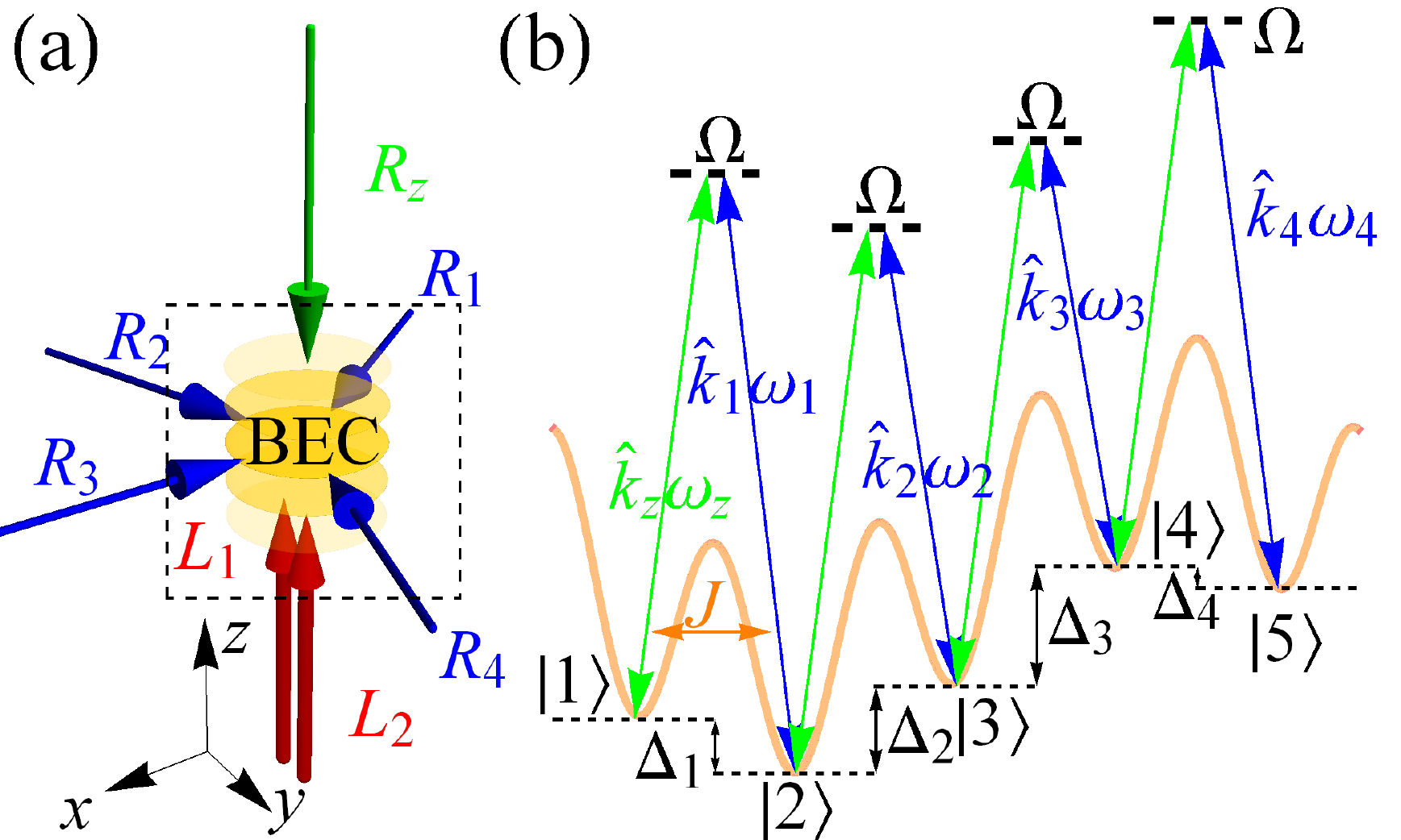}
\caption{(Color online) (a) Proposed experimental scheme for generating
superfluid-quasicrystals with 5-fold rotational symmetry. The superlattice
is generated by two optical lattices with different periods. A potential
gradient can be generated using a magnetic field gradient. One Raman beam $%
R_{z}$ in the $z$-direction and four others ($R_{j},j=1,2,3,4$) in the $x$-$%
y $ plane generate the Raman coupling between neighboring wells. (b) SOC in
one unit cell. Each Raman process only couples well $j$ to its adjacent
neighbor $j+1$. The hopping between neighboring unit cells is irrelevant
because of the large bias between $\left\vert 1\right\rangle $ and $%
\left\vert 5\right\rangle $. Only Raman-assisted inter-well tunneling in a
unit cell is considered.}
\label{fig1}
\end{figure}

\emph{Experimental scheme and Hamiltonian}. We consider a 3D BEC confined in
a tilted superlattice potential
\begin{equation}
V_{SL}(z)=V_{1}\sin ^{2}(k_{L1}z)+V_{2}\sin ^{2}(k_{L2}z+\phi _{12})+\alpha
_{z}z
\end{equation}%
along the $z$-direction [Fig.~\ref{fig1}(a)] with $k_{L2}=k_{L1}/5$. Here
two lattices can come from the same laser source with the second lattice
potential formed by two beams intersecting with an angle $\theta =2\arcsin
(1/5)\approx 23^{\circ }$. The linear potential $\alpha _{z}z$ can be
realized with a magnetic field gradient. Note that this superlattice does
not defy the definition for superfluid-quasicrystal because it only breaks
the translational symmetry in the $z$-direction, while the spontaneous
formation of (quasi)crystal order is on the $x$-$y$ plane. We denote five
wells in each unit cell as five pseudospins and the effective couplings $%
\Omega $ between neighboring spins are induced by 5 Raman beams, one in the $%
z$ direction and four in the $x$-$y$ plane with designated wavevectors $\hat{%
k}_{j} $ [Fig.~\ref{fig1}(b)]. We choose suitable parameters $\phi _{12}$, $%
\alpha _{z}$ such that the energy bias between neighboring wells $|\Delta
_{j}|\gg J $ to avoid direct hopping ($J$ is the bare tunneling rate without
Raman coupling) and $||\Delta _{j}|-|\Delta _{i}||\gg \Omega $ for $i\neq j$
so that two neighboring wells are coupled only by one specific Raman pair.

\begin{figure}[t]
\centering
\includegraphics[width=0.48\textwidth]{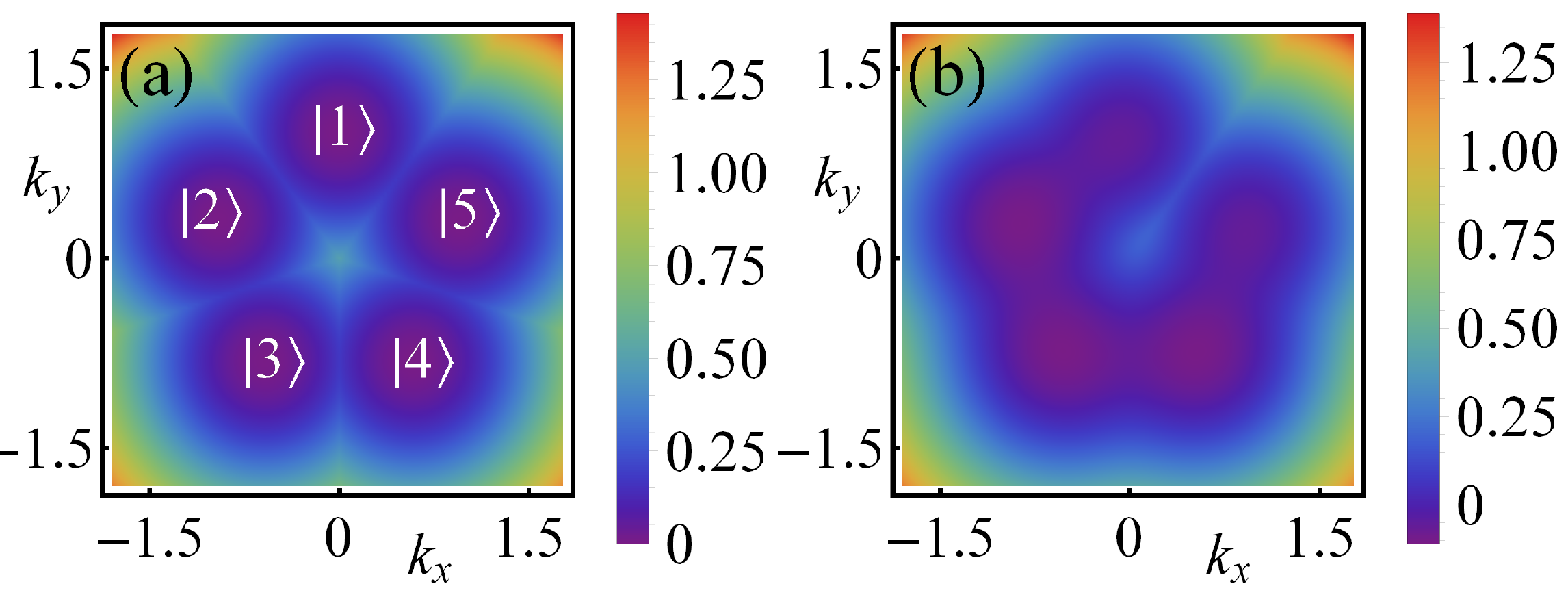}
\caption{(Color online) Single particle lowest band dispersion. (a) $\Omega
=0$. The minima form a regular pentagon structure in the momentum space. (b)
$\Omega =0.4$. The minima are strongly coupled and all spin components are
mixed. Wells $1$ and $5$ are uncoupled and a barrier between them can be
clearly observed. The five minima are axisymmetric to $\hat{p}_{3}$ for any $%
\Omega $.}
\label{fig2}
\end{figure}

The effective single particle Hamiltonian $H_{0}$ in the $x$-$y$ plane can
be written as%
\begin{equation}
H_{0}=\sum_{j=1}^{5}\left[ \frac{(\hat{p}-\hat{p}_{j})^{2}}{2}+\delta _{j}%
\right] |j\rangle \langle j|+\sum_{j=1}^{4}\left( \frac{\Omega }{2}|j\rangle
\langle j\text{+}1|+h.c.\right) ,
\end{equation}%
after a standard unitary transformation of the pseudospin phases to remove
the spatial dependence of the Raman coupling \cite{MyApp}. Here we choose
the units as $\hbar =k_{R}=m=1$, where $k_{R}$ is recoil wavevector and $m$
is atomic mass. The energy unit $E_{R}=\hbar ^{2}k_{R}^{2}/m=1$. $\delta
_{j} $ is the detuning determined by the Raman transition. $\hat{p}_{j}$
satisfying $\hat{p}_{j}=\hat{p}_{j-1}+2\hat{k}_{j-1}$ and $\hat{p}_{1}=-%
\frac{2}{5}\sum_{j=1}^{5}(5-j)\hat{k}_{j}$ \cite{MyApp}. In order to
generate good superfluid-quasicrystals, we consider a regular pentagon (all
minima form an equilateral polygon) with $\hat{p}_{1}=(0,1)$, $\hat{p}_{2}=(-%
\sqrt{5/8+\sqrt{5}/8}$,$~(-1+\sqrt{5})/4)$, $\hat{p}_{3}=(-\sqrt{5/8-\sqrt{5}%
/8},~(-1-\sqrt{5})/4)$, $\hat{p}_{4}=(\sqrt{5/8-\sqrt{5}/8},~(-1-\sqrt{5}%
)/4) $, and $\hat{p}_{5}=(\sqrt{5/8+\sqrt{5}/8},~(-1+\sqrt{5})/4)$ [see Fig.%
\ref{fig2}(a)], which can be realized using $\hat{k}_{j}=\left( \hat{p}%
_{j+1}-\hat{p}_{j}\right) /2$ ($j=1,2,3,4$) for four Raman lasers in the $xy$
plane.

For $\Omega =0$ and $\delta _{j}=0$ [Fig.~\ref{fig2}(a)], five minima
distribute over the vertexes of a regular pentagon. All spin components are
uncoupled and only occupy one minimum. As a strong Raman coupling $\Omega
=0.4$ is ramped on [Fig.~\ref{fig2}(b)], the minima are coupled as an open
boundary chain without coupling between head and tail. The spin components
are mixed at each minimum and each well starts to merge with its adjacent
neighbor. Because the locations of all minima [labelled as in Fig.~\ref{fig2}%
(a)] are axisymmetric to the vector $\hat{p}_{3}$, the two uncoupled minima
(the head and the tail) disappear first at certain critical value of $\Omega
$ (we label the remaining minima as $2$ to $4$). Finally, the remaining
three minima merge into one (minimum $3$) approximately located at $%
(-0.207,-0.286)$ \cite{MyApp} when $\Omega $ is extremely strong and its
location is still along the same line as $\hat{p}_{3}$.

\emph{Phase diagram}. We now study new quantum phases emerging from
interactions between atoms, which can be described by the GPE under the
mean-field approximation with energy density
\begin{equation}
\epsilon =\int \frac{d\hat{r}}{V}\left[ \psi ^{\dagger }H_{0}\psi +\frac{%
c_{0}}{2}n_{i}^{2}+\frac{c_{2}}{2}\sum\nolimits_{i=1}^{4}n_{i}n_{i+1}\right]
,
\end{equation}%
where $\psi $ is the 5-component spinor wavefunction, $n_{i}=\psi
_{i}^{\dagger}\psi _{i}$ is the density for the spin component $i$, $c_{0}$
and $c_{2}\approx \frac{J}{\Delta }c_{0}$ are density interaction for the
same and neighboring spins, respectively. For realistic parameters, $%
J/\Delta \sim 1/20$, the neighboring spin interaction can be ignored \cite%
{Li2017} and this is a crucial condition for realizing
superfluid-quasicrystals or supersolids in experiments. The wavefunction is
normalized by the average atomic density as $V^{-1}\int d\hat{r}\psi
^{\dagger }\psi =\bar{n}$ with $V$ being the system volume. We obtain the
ground state using both variational ansatz analysis and direct numerical
simulation of the GPE, and they agree well.

The general form of the variational ansatz is
\begin{equation}
\psi =\sqrt{\bar{n}}\sum\nolimits_{j=1}^{5}C_{j}e^{i\hat{k}_{m,j}\cdot \hat{r%
}}\xi _{j},
\end{equation}%
where $C_{j}$ are complex numbers satisfying normalization relation $%
\sum_{j}|C_{j}|^{2}=1$, $\hat{k}_{m,j}$ denotes each minimum in momentum
space and $\xi _{j}$ are the spinor part of wavefunction $\xi _{j}=(\cos
\alpha _{j}\cos \beta _{j}\cos \gamma _{j}$, $\cos \alpha _{j}\cos \beta
_{j}\sin \gamma _{j}$,$~\sin \alpha _{j}$, $\cos \alpha _{j}\sin \beta
_{j}\sin \eta _{j}$, $\cos \alpha _{j}\sin \beta _{j}\cos \eta _{j})^{T}$.
We assume $\xi _{1,j}=\xi _{5,6-j}$, $\xi _{2,j}=\xi _{4,6-j}$ ($\xi _{i,j}$
stands for the $j$th component of spinor $\xi _{i}$), and $\hat{k}_{m,1}(%
\hat{k}_{m,2})$ and $\hat{k}_{m,5}(\hat{k}_{m,4})$ are axisymmetric to
vector $\hat{k}_{m,3}$ based on the symmetry of the Hamiltonian. Generally
it is challenging to optimize the energy density functional with so many
variables. However, in the weak interaction region $\bar{n}c_{0}\ll 1$, the
BEC wavefunction at each band minimum is quite close to the single particle
spinor wavefunction, which can thus be used to fix $\xi _{j}$ for the
variational calculation. Similar method was used previously for studying
spin-1 spin-orbit coupled BEC, which gives all phases as those in full
variational calculation, although the phase boundary may be slightly
different for stronger interaction \cite{Luo2017}. We also find that $%
\left\vert C_{1}\right\vert =\left\vert C_{5}\right\vert $ and $\left\vert
C_{2}\right\vert =\left\vert C_{4}\right\vert $ hold in weak interaction
cases. The ground state energy is degenerate with respect to relative phases
between $C_{j}$ and the system spontaneously chooses one set of relative
phases for the superfluid-quasicrystal and supersolid stripe phases.

\begin{figure}[t]
\centering
\includegraphics[width=0.48\textwidth]{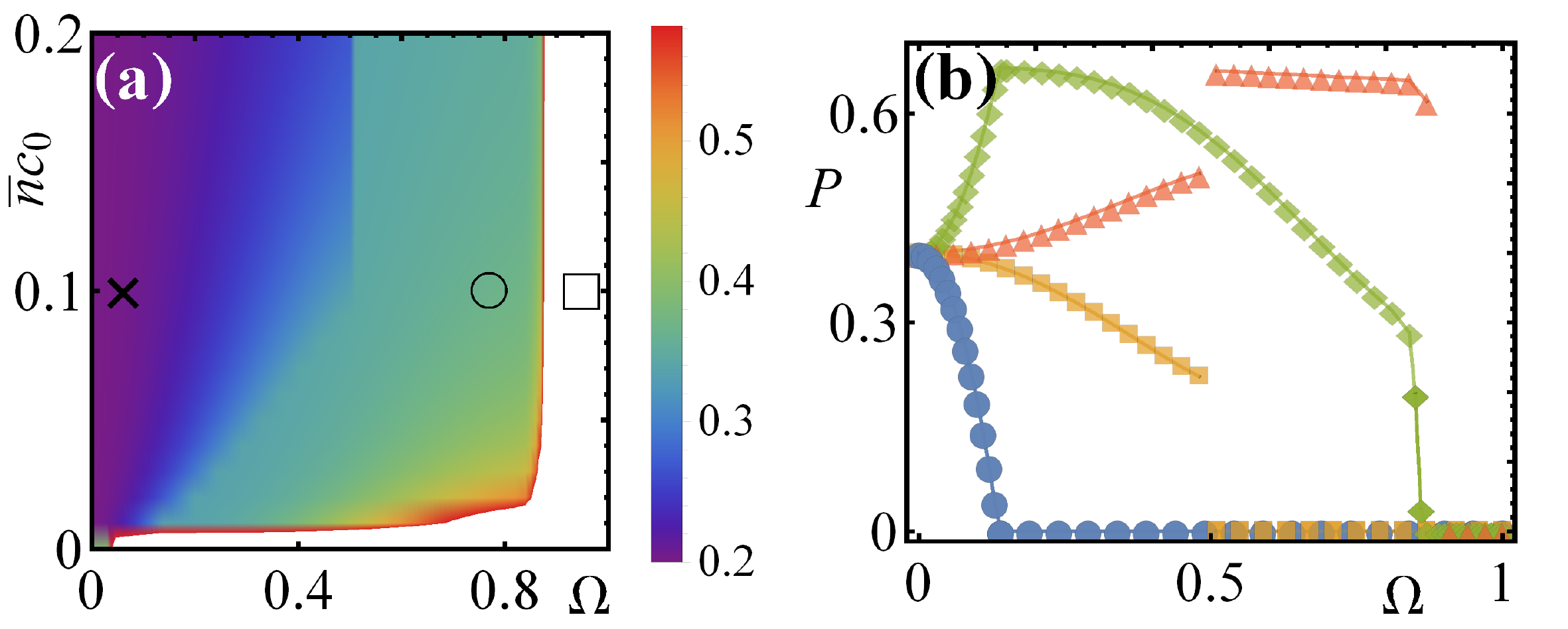}
\caption{(Color online) (a) Phase diagram from the variational ansatz
analysis when all the detunings $\protect\delta_j$ are set to be zero. The
color represents the value of $|C_{3}|^{2}$. The white region is the
plane-wave phase. The three symbols (cross, circle, and square stand for $%
\Omega =0.06$, $0.77$ and $0.95 $) along $\bar{n}c_{0}=0.1$ are examples for
superfluid-quasicrystal, supersolid, and plane-wave phases. (b) Phase
transitions between different phases. The blue circles (green rhombus) and
orange squares (red triangles) show how $|C_{1}|^{2}+|C_{5}|^{2}$ ($%
|C_{2}|^{2}+|C_{4}|^{2}$) varies with Raman coupling for $\bar{n}c_{0}=0.01$
and $0.2$, respectively.}
\label{fig3}
\end{figure}

In Fig.~\ref{fig3}(a), we plot the phase diagram with respect to the
interaction strength $\bar{n}c_{0}$ and Raman coupling strength $\Omega $
obtained from the variational ansatz calculation, where the color shows the
occupied probability $|C_{3}|^{2}$ at the momentum minimum 3. At a finite $%
\Omega $, 3 has the lowest energy, therefore atoms only occupy 3 without
interaction, leading to a plane-wave phase. On the other hand, a strong
density-density interaction prefers the equal occupation of all minima.
Therefore the competition between Raman coupling and interaction may render
different phases, as shown in Fig.~\ref{fig3}(a).

\begin{figure}[t]
\centering
\includegraphics[width=0.48\textwidth]{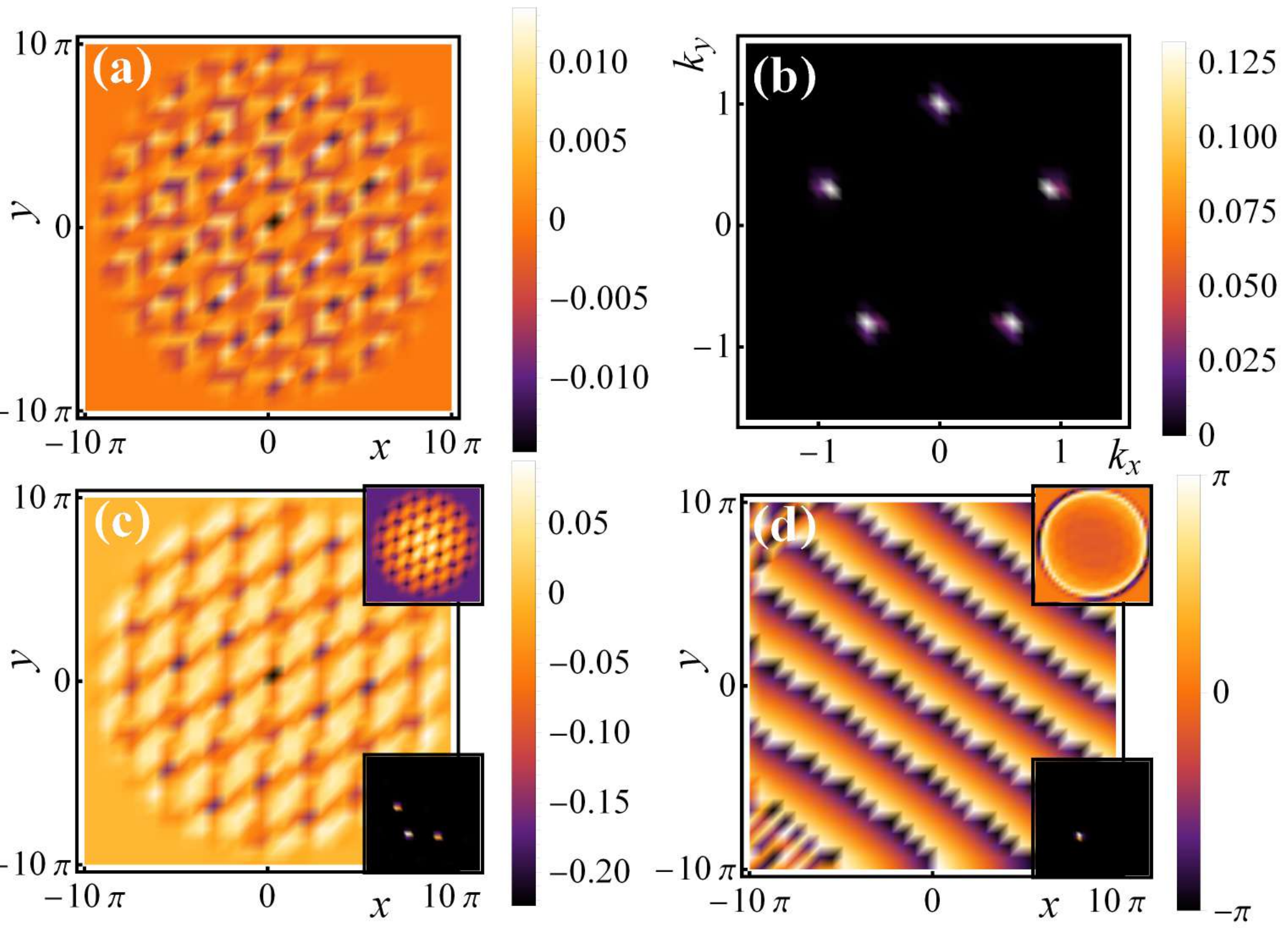}
\caption{(Color online) (a)[(b)] Real-space (momentum-space) distribution
for the superfluid-quasicrystal phase from GPE simulation. The parameters
are the same as the black cross marked in Fig.~\protect\ref{fig3}(a). The
five minima in momentum space are evenly populated. (c) Spatial distribution
for a supersolid phase [parameters chosen as black circle in Fig.~\protect
\ref{fig3}(a)] with three minima populated. The top and bottom insets show
the real-space density of state $|3\rangle $ and the momentum-space
distribution respectively. (d) Phase distribution for state $|3\rangle $ at
a nonzero momentum plane-wave phase, and the insets give real (top) and
momentum (bottom) space distributions. The parameters correspond to the
black square in Fig.~\protect\ref{fig3}(a). All real-space density
distributions (including following panels) are obtained through subtracting
the real-space density at zero Raman coupling from those of finite $\Omega $
to rule out the large density variation across the harmonic trap. The plots
are rescaled over the average density in the trap. The total momentum-space
distribution is the direct summation of that for each pseudospin component.}
\label{fig4}
\end{figure}

In the small $\Omega $ region, all five minima are equally populated with
the same probability $1/5$ due to interaction, forming a
superfluid-quasicrystal [cross in \ref{fig3}(a)]. This new quantum matter is
confirmed by the real and momentum space density distributions [Fig.~\ref%
{fig4}(a,b)] obtained from the GPE simulation in a harmonic trap. We see the
distribution in the real space is indeed in lack of translational symmetry,
while in momentum space five equally populated peaks form a regular pentagon
with each vertex designated as $\hat{p}_{j}$, showing the 5-fold rotational
symmetry of the superfluid-quasicrystal phase. Note that the harmonic trap
breaks the degeneracy of the ground state and fixes the relative phases
between $C_{j}$.

With the increase in $\Omega $, the occupation of five minima changes to
three, leading to a supersolid stripe phase [circle in Fig. \ref{fig3}(a)],
where $|C_{3}|^{2}$ increases to $\sim 0.35$. The resulting real and
momentum space density distributions from GPE are shown in Fig.~\ref{fig4}%
(c). Here a clear translational symmetry in the real space is observed. In
the momentum space distribution (bottom inset), three minima are occupied
unevenly and minimum 3 has a larger weight. Here $\Omega $ is quite large
and the spin components in each minimum are mixed. Consequently, the spatial
distribution of state $|3\rangle $ exhibits clear density modulation (top
inset). \ For a very large $\Omega $, all minima merge to one and the system
enters a plane-wave phase [square in \ref{fig3}(a)]. In Fig.~\ref{fig4}(d),
we plot its phase distribution obtained from GPE, which shows a stripe
pattern as expected. The overall real space density distribution (top inset)
is a Gaussian-type wavepacket and the BEC occupies one point in the momentum
space (bottom inset).

We characterize the transition between these phases in Fig.~\ref{fig3}(b),
where we plot the populations $P_{15}=|C_{1}|^{2}+|C_{5}|^{2}$ and $%
P_{24}=|C_{2}|^{2}+|C_{4}|^{2}$ with respect to $\Omega $ for two different
interaction strengths $\bar{n}c_{0}=0.01$ and 0.2. In the weak interaction
case $\bar{n}c_{0}=0.01$, $P_{15}$ smoothly decreases to $0$ and $P_{24}$
has a sharp turn at certain $\Omega $, showing a second-order phase
transition from superfluid-quasicrystal to supersolids. This occurs when the
energy bias between minima 1 and 2 with increasing $\Omega $ is larger than
the interaction energy cost. For the strong interaction $\bar{n}c_{0}=0.2$, $%
P_{15}$ ($P_{24}$) shows a sudden drop at $\Omega \sim 0.5$, showing a
first-order phase transition at the point where single particle five minima
merge into three. Before the transition, the interaction energy cost is so
strong that $P_{15}$ is always nonzero. Around $\Omega \sim 0.8$, the three
minima merge into one and $P_{24}$ also suddenly drops to zero, showing a
first order phase transition to the plane wave phase.

\begin{figure}[t]
\centering
\includegraphics[width=0.48\textwidth]{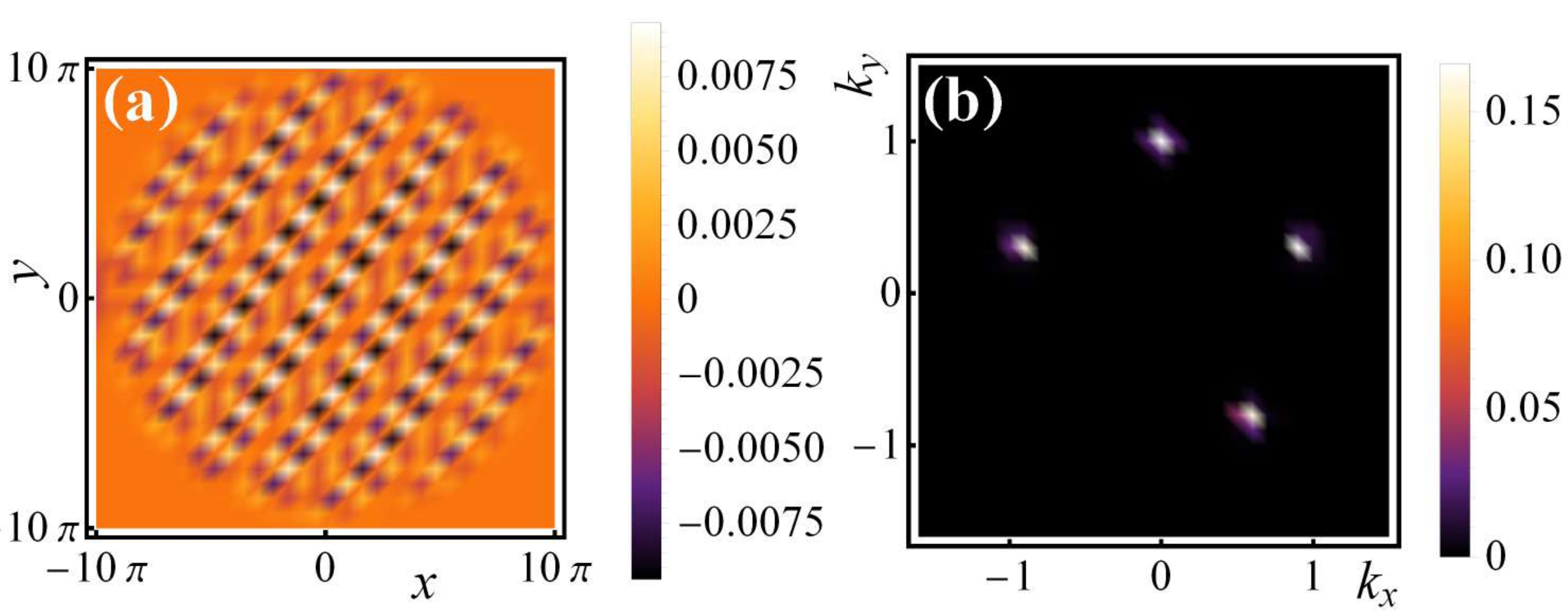}
\caption{(Color online) (a)[(b)] Real-space (momentum-space) distribution
for a supersolid phase when one minimum is knocked off with a small detuning
$\protect\delta _{3}\sim 0.01E_{R}$ in the superfluid-quasicrystal phase in
Fig.~\protect\ref{fig4}(a,b).}
\label{fig5}
\end{figure}

In addition to varying $\Omega$, we may also adjust detuning $\delta_j$ to
change the relative population of the minima. In Fig.~\ref{fig5}(a,b), we
plot the real and momentum-space distribution from GPE by adding a small
detuning $\delta_3 \sim 0.01E_{R}$ in spin state $|3\rangle $ for the
superfluid-quasicrystal state in Fig~\ref{fig4}(a). We see that the minimum
3 is now knocked out and the translational symmetry is restored. The BEC
becomes a supersolid with exotic real-space distribution because it is still
populated on four of five vertexes of a regular pentagon. The interplay
among Raman coupling, detuning, and interaction leads to a rich phase
diagram and hence enables the designing and engineering of new
superfluid-quasicrystal and supersolid phases.

\emph{Experimental realization and detection}. The experimental realization
of our scheme is in the same spirit as recent experimental reports on
observing supersolid stripe phases~\cite{Li2017,Li2016}. Consider $N=10^{5}$
${}^{23}$Na atoms confined in a super-lattice with five wells in one unit
cell and the condensate is initially split into each well equally with an
average density around $\bar{n}=0.5\times 10^{14}~\text{cm}^{-3}$. We choose
$k_{R}=1064$ nm and thus, $E_{R}=7672$ Hz, for Raman lasers. The Raman
coupling strength $\Omega $ can be tuned as low as $300$ Hz \cite%
{Li2017,Li2016}, that is, $\Omega \lesssim 0.08E_{R}$, which well resides in
the superfluid-quasicrystal region. In real experiments, $%
(c_{0}-c_{2})/c_{0}\approx 1$, therefore the neighboring-spin interaction
term can be neglected. The same spin density interaction strength can be
evaluated with $c_{0}=4\pi \hbar ^{2}a_{s}/m$, where $a_{s}$ is the two-body
scattering length. Taking $a_{s}=50a_{0}$ ($a_{0}$ is Bohr radius) \cite%
{Abeelen1999,Knoop2011}, we have $\bar{n}c_{0}\approx 0.1E_{R}$, which is
sufficiently strong for the observation of superfluid-quasicrystals,
although a larger density $\bar{n}$ or scattering length $a_{s}$ (tuned by
Feshbach resonance) yields a larger parameter region (Fig. \ref{fig3}) and
is better for the observation of superfluid-quasicrystal phases.

In experiments, the superfluid-quasicrystal phases may be observed in the
time-of-flight (TOF) image, where five equally populated peaks are formed in
the momentum space at designated positions [Fig.~\ref{fig4}(b)]. By
measuring $P_{15}$ and $P_{24}$ in TOF, the quantum phase transition between
different phases in Fig. \ref{fig3}(b) can be detected. Another way to
observe the superfluid-quasicrystal phases is using Bragg scattering,
similar to that for supersolids \cite{Li2017}, where the Bragg diffraction
patterns for superfluid-quasicrystals should give peaks possessing 5-fold
rotational symmetry~\cite{Levine1984}.

\emph{Discussion and Conclusion}. Our proposed experimental setup can be
straightforwardly generalized for realizing other superfluid-quasicrystal
phases with higher order rotational symmetry, such as $n=7$, where 7 Raman
lasers are needed with a similar experimental setup for 7 wells in a
superlattice. Similar idea can also be applied to generate
superfluid-quasicrystal phases in a spin-orbit coupled BEC with atomic
hyperfine state pseudospins \cite%
{Lin2011,Zhang2012b,Qu2013a,Olson2014,Hamner2014,Wang2012,Cheuk2012,Williams2013,Lev,Jo,Huang2016, Meng2016,Pan2016}%
, where supersolid stripe phases have been proposed \cite{Stanescu2008,
Wu2011, wang2010spin, ho2011bose, li2012quantum, zhang2012mean, hu2012spin,
ozawa2012stability, sun2016interacting, yu2016phase,
martone2016tricriticalities} for both 1D and 2D SOC, but have not been
observed in experiments. Note that our proposed scheme for
superfluid-quasicrystals requires five almost degenerate band minima for
five different spins to generate a regular pentagon in the momentum space.
In experiments, an effective 2D SOC (not exactly Rashba) has been
experimentally realized recently \cite{Huang2016, Meng2016,Pan2016} by
coupling three spin states at three degenerate band minima, although the
resulting band minimum path in the lowest band is not a flat ring as
expected from a Rashba SOC. In our scheme, no flat Rashba ring is needed and
5 Raman lasers with suitable wavevectors and polarizations are chosen such
that the effective band minima are formed at $\hat{p}_{j}$ for a regular
pentagon. The crucial difficulty comes from the interaction that is almost
isotropic between any spin states. This difficulty may be resolved using $%
^{133}$Cs atoms \cite{Chin2015}, where the interaction may be tuned by
Feshbach resonance to favor the equal occupation of five minima in the
momentum space, instead of the plane-wave at one minimum.

In summary, we have proposed a scheme for realizing superfluid-quasicrystal
stripe phases using a BEC in a 1D quintuple-well optical superlattice with
Raman-assisted tunneling. Through variational and GPE analysis, we show
there is a rich phase diagram containing superfluid-quasicrystals,
supersolids, plane-wave phases, and their phase transitions. Our proposed
experimental setup should lay out a platform for future theoretical and
experimental investigations of such exotic novel quantum matter.

\begin{acknowledgments}
\textbf{Acknowledgments}: This work is supported by AFOSR
(FA9550-16-1-0387), NSF (PHY-1505496), and ARO (W911NF-17-1-0128).
\end{acknowledgments}


\newpage \clearpage
\onecolumngrid
\appendix

\section{Supplementary materials}

\subsection{Single particle Hamiltonian for spin-orbit-coupling}

We derive the effective single-particle Hamiltonian for our pseudospin
system with $n=5$ and the method can be generalized to arbitrary $n$.
Results for $n=2$ have been studied with great details in Ref.~\cite{Li2016}%
. The single particle Hamiltonian consists of two parts: the superlattice
potential $V_{SL}$ in the $z$-direction and the Raman coupling.

Consider a superlattice with a tilted potential along the $z$ direction
\begin{equation}
H_{SL}=\frac{\hat{p}^{2}}{2}+\frac{p_{z}^{2}}{2}+V_{1}\sin
^{2}(k_{L1}z)+V_{2}\sin ^{2}(k_{L2}z+\phi _{12})+\alpha _{z}z,
\end{equation}%
where $k_{L1}=\pi /d$, $k_{L2}=5\pi /d$, and $d$ is the period for the long
lattice. In the tight-binding limit,

\begin{eqnarray*}
H_{SL} &=&\frac{\hat{p}^{2}}{2}+\left( \Delta _{1}+\Delta _{2}\right)
\sum_{m} |\Psi _{1,m}\rangle \langle \Psi _{1,m}| +\Delta _{2}\sum_{m}|\Psi
_{2,m}\rangle \langle \Psi _{2,m}| \\
&&-\Delta _{3}\sum_{m}|\Psi _{4,m}\rangle \langle \Psi _{4,m}|-\left( \Delta
_{4}+\Delta _{3}\right) \sum_{m}|\Psi _{5,m}\rangle \langle \Psi _{5,m}| \\
&&+J\sum_{m}\sum_{j=1}^{4}\left( |\Psi _{j,m}\rangle \langle \Psi _{j+1,m}|+
\mathrm{{h.c.} }\right) +\sum_{m}\sum_{l=1}^{5}\sum_{j=1}^{5}\left(
J_{l,j}|\Psi _{l,m}\rangle \langle \Psi _{j,m+1}|+ \mathrm{h.c.}\right) ,
\end{eqnarray*}%
where $|\Psi _{j,m}\rangle $ is the onsite wavefunction of well $j$ in the $%
m $-th unit cell. Well $3$ is set as the reference of zero-energy and $%
\Delta _{j}$ is the energy difference between wells $j$ and $j+1$. Hereafter
we neglect the coupling $J_{l,j}$ between adjacent unit cells because there
is no Raman assisted tunneling between wells in different unit cells. Since $%
J\ll \Delta _{j}$, the wavefunctions can be expanded to the first-order
\begin{eqnarray}
|1,m\rangle &=&|\Psi _{1,m}\rangle +\frac{J}{\Delta _{1}}|\Psi _{2,m}\rangle
,~|2,m\rangle =|\Psi _{2,m}\rangle -\frac{J}{\Delta _{1}}|\Psi _{1,m}\rangle
+\frac{J}{\Delta _{2}}|\Psi _{3,m}\rangle , \\
|3,m\rangle &=&|\Psi _{3,m}\rangle -\frac{J}{\Delta _{2}}|\Psi _{2,m}\rangle
+\frac{J}{\Delta _{3}}|\Psi _{4,m}\rangle ,~|4,m\rangle =|\Psi _{4,m}\rangle
-\frac{J}{\Delta _{3}}|\Psi _{3,m}\rangle +\frac{J}{\Delta _{4}}|\Psi
_{5,m}\rangle ,~|5,m\rangle =|\Psi _{5,m}\rangle -\frac{J}{\Delta _{4}}|\Psi
_{4,m}\rangle .  \nonumber
\end{eqnarray}

The four Raman couplings are
\begin{equation}
V_{Raman,j}=\Omega _{j}\cos \left( k_{z}z+\hat{k}_{j}\cdot \hat{r}-\delta
_{R,j}t\right) ,
\end{equation}%
where $k_{z}=5\pi /(2d)$. Expanding $H_{SL}$ under above perturbed basis, we
obtain
\begin{eqnarray}
H_{SL} &=&\frac{\hat{p}^{2}}{2}+\left( \Delta _{1}+\Delta _{2}\right)
\sum_{m}\left( |1,m\rangle \langle 1,m|\right) +\Delta
_{2}\sum_{m}|2,m\rangle \langle 2,m| \\
&&-\left( \Delta _{4}+\Delta _{3}\right) \sum_{m}|5,m\rangle \langle
5,m|-\Delta _{3}\sum_{m}|4,m\rangle \langle 4,m|  \nonumber \\
&+&\sum_{\hat{p},\hat{p}^{\prime }}|\hat{p}\rangle \left( \sum_{n^{\prime
}}\sum_{l=1}^{5}\sum_{j=1}^{5}|l,m\rangle \langle l,m|\langle \hat{p}|\Omega
_{j}\cos \left( k_{z}z+\hat{k_{j}}\cdot \hat{r}-\delta _{j}t\right) |\hat{p}%
^{\prime }\rangle |j,m\rangle \langle j,m|\right) \langle \hat{p}^{\prime }|,
\nonumber
\end{eqnarray}%
where
\begin{eqnarray}
\langle 1,m|\cos \left( k_{z}(z-z_{m})\right) |2,m\rangle &=&-\frac{J}{%
\Delta _{1}},~\langle 2,m|\cos \left( k_{z}(z-z_{m})\right) |3,m\rangle =-%
\frac{J}{\Delta _{2}}, \\
\langle 3,m|\cos \left( k_{z}(z-z_{m})\right) |4,m\rangle &=&\frac{J}{\Delta
_{3}},~\langle 4,m|\cos \left( k_{z}(z-z_{m})\right) |5,m\rangle =\frac{J}{%
\Delta _{4}},  \nonumber \\
\langle 1,m|\sin \left( k_{z}(z-z_{m})\right) |2,m\rangle &=&\frac{J}{\Delta
_{1}},~\langle 2,m|\sin \left( k_{z}(z-z_{m})\right) |3,m\rangle =-\frac{J}{%
\Delta _{2}},  \nonumber \\
\langle 3,m|\sin \left( k_{z}(z-z_{m})\right) |4,m\rangle &=&-\frac{J}{%
\Delta _{3}},~\langle 4,m|\sin \left( k_{z}(z-z_{m})\right) |5,m\rangle =%
\frac{J}{\Delta _{4}},  \nonumber \\
\langle j,m|\cos \left( k_{z}(z-z_{m})\right) |j,m\rangle &=&\sin (\frac{\pi
}{2}j),~\langle j,m|\sin \left( k_{z}(z-z_{m})\right) |j,m\rangle =-\cos (%
\frac{\pi }{2}j),  \nonumber
\end{eqnarray}%
up to the first order, $z_{m}=md$ is the position of the $1$-st well in the $%
m$-th unit cell. With these relations, the Raman potential can be
reformulated as
\begin{eqnarray}
&&\sum_{l=1}^{5}\sum_{j=1}^{5}|l,m\rangle \langle l,m|\langle \hat{p}|\Omega
\cos \left( k_{z}z+\hat{k_{j}}\cdot \hat{r}-\delta _{j}t\right) |\hat{p}%
^{\prime }\rangle |j,m\rangle \langle j,m|  \label{RamanTerms} \\
&=&~\Omega _{j}\cos \phi _{m,j}\left( |1,m\rangle \langle 1,m|-|3,m\rangle
\langle 3,m|+|5,m\rangle \langle 5,m|\right)  \nonumber \\
&&+\Omega _{j}\cos \phi _{m,j}\left( -\frac{J}{\Delta _{1}}|1,m\rangle
\langle 2,m|-\frac{J}{\Delta _{2}}|2,m\rangle \langle 3,m|+\frac{J}{\Delta
_{3}}|3,m\rangle \langle 4,m|+\frac{J}{\Delta _{4}}|4,m\rangle \langle
5,m|\right)  \nonumber \\
&&+\Omega _{j}\sin \phi _{m,j}\left( |2,m\rangle \langle 2,m|-|4,m\rangle
\langle 4,m|\right)  \nonumber \\
&&+\Omega _{j}\sin \phi _{m,j}\left( \frac{J}{\Delta _{1}}|1,m\rangle
\langle 2,m|-\frac{J}{\Delta _{2}}|2,m\rangle \langle 3,m|-\frac{J}{\Delta
_{3}}|3,m\rangle \langle 4,m|+\frac{J}{\Delta _{4}}|4,m\rangle \langle
5,m|\right) ,  \nonumber
\end{eqnarray}%
where $\phi _{m,j}=\pi m/2+\hat{k}_{j}\cdot \hat{r}-\delta _{R,j}t$. $\delta
_{R,j}$ is chosen to be close to $\Delta _{j}$, but off-resonate to other $%
\Delta _{i},i\neq j$ so that the Raman potential $V_{Raman,j}$ only couples $%
|j,m\rangle $ to its neighbor state $|j+1,m\rangle $.

Without Raman coupling, atoms in well $j$ are not coupled with other wells,
yielding a state at $q=(j-1)2\pi /(5d)$ of the lowest band of the
superlattice
\begin{equation}
|\psi _{q=(j-1)2\pi /(5d)}^{(j)}\rangle =\sum_{m=1}^{N}\frac{1}{\sqrt{N}}e^{i%
\frac{2\pi (j-1)}{5d}\left[ z_{m}+(j-1)\frac{d}{5}\right] }|j,m\rangle .
\end{equation}%
Here $N$ is the number of unit cells in the superlattice.

The intra-band couplings
\begin{eqnarray}
\langle \psi _{q=j2\pi /(5d)}^{(j)}|V_{Raman,j}|\psi _{q=(j-1)2\pi
/(5d)}^{(j)}\rangle &=&\sum_{m,m^{\prime }}\frac{1}{N}e^{i\frac{2\pi }{5}%
\left[ (m^{\prime }-m-\frac{1}{5})j-m^{\prime }+\frac{1}{5}\right] }\langle
j,m|V_{Raman,j}|j,m^{\prime }\rangle \\
&=&\sum_{m}\frac{1}{N}e^{-i\frac{2\pi }{5}\left[ m+(j-1)\frac{1}{5}\right]
}\langle j,m|V_{Raman,j}|j,m\rangle ,  \nonumber \\
\langle \psi _{q=(j-1)2\pi /(5d)}^{(j+1)}|V_{Raman,j}|\psi _{q=j2\pi
/(5d)}^{(j+1)}\rangle &=&\sum_{m,m^{\prime }}\frac{1}{N}e^{i\frac{2\pi }{5}%
\left[ (m^{\prime }-m-\frac{1}{5})j-m\right] }\langle
j+1,m|V_{Raman,j}|j+1,m^{\prime }\rangle \\
&=&\sum_{m}\frac{1}{N}e^{-i\frac{2\pi }{5}\left[ m-\frac{j}{5}\right]
}\langle j+1,m|V_{Raman,j}|j+1,m\rangle ,  \nonumber
\end{eqnarray}%
and the SOC
\begin{eqnarray}
\langle \psi _{q=j2\pi /(5d)}^{(j+1)}|V_{Raman,j}|\psi _{q=(j-1)2\pi
/(5d)}^{(j)}\rangle &=&\sum_{m,m^{\prime }}\frac{1}{N}e^{i\frac{2\pi }{5}%
\left[ (m^{\prime }-m-\frac{2}{5})j-m^{\prime }+\frac{1}{5}\right] }\langle
j+1,m|V_{Raman,j}|j,m^{\prime }\rangle  \label{SOCGen} \\
&=&\sum_{m}\frac{1}{N}e^{-i\frac{2\pi }{5}\left[ m+(2j-1)\frac{1}{5}\right]
}\langle j+1,m|V_{Raman,j}|j,m\rangle .  \nonumber
\end{eqnarray}

The intra-band terms cause a density modulation. If only near-resonant terms
are kept, the single particle Hamiltonian can be written as
\[
\left(
\begin{array}{ccccc}
\frac{\hat{p}^{2}}{2}+\Delta _{1}+\Delta _{2} & c_{p,1}\frac{J}{\Delta _{1}}%
\Omega _{1}e^{-i(\hat{k}_{1}\cdot \hat{r}-\delta _{R,1}t)} & 0 & 0 & 0 \\
c_{p,1}^{\ast }\frac{J}{\Delta _{1}}\Omega _{1}e^{i(\hat{k}_{1}\cdot \hat{r}%
-\delta _{R,1}t)} & \frac{\hat{p}^{2}}{2}+\Delta _{2} & c_{p,2}\frac{J}{%
\Delta _{2}}\Omega _{2}e^{-i(\hat{k}_{2}\cdot \hat{r}-\delta _{R,2}t)} & 0 &
0 \\
0 & c_{p,2}^{\ast }\frac{J}{\Delta _{2}}\Omega _{2}e^{i(\hat{k}_{2}\cdot
\hat{r}-\delta _{R,2}t)} & \frac{\hat{p}^{2}}{2} & c_{p,3}\frac{J}{\Delta
_{3}}\Omega _{3}e^{-i(\hat{k}_{3}\cdot \hat{r}-\delta _{R,3}t)} & 0 \\
0 & 0 & c_{p,3}^{\ast }\frac{J}{\Delta _{3}}\Omega _{3}e^{i(\hat{k}_{3}\cdot
\hat{r}-\delta _{R,3}t)} & \frac{\hat{p}^{2}}{2}-\Delta _{3} & c_{p,4}\frac{J%
}{\Delta _{4}}\Omega _{4}e^{-i(\hat{k}_{4}\cdot \hat{r}-\delta _{R,4}t)} \\
0 & 0 & 0 & c_{p,4}^{\ast }\frac{J}{\Delta _{4}}\Omega _{4}e^{i(\hat{k}%
_{4}\cdot \hat{r}-\delta _{R,4}t)} & \frac{\hat{p}^{2}}{2}-\Delta
_{3}-\Delta _{4}%
\end{array}%
\right) ,
\]%
where complex constants $c_{p,j}$ for each SOC terms are determined by Eqs.~(%
\ref{RamanTerms}) and (\ref{SOCGen}). A unitary transformation $U_{t}$
defined as $|1\rangle _{t}\rightarrow |1\rangle _{t}e^{-i(\delta
_{R,1}+\delta _{R,2})t}$, $|2\rangle _{t}\rightarrow |2\rangle
_{t}e^{-i\delta _{R,2}t}$, $|3\rangle _{t}\rightarrow |3\rangle _{t}$, $%
|4\rangle _{t}\rightarrow |4\rangle _{t}e^{-i\delta _{R,3}t}$, $|5\rangle
_{t}\rightarrow |5\rangle _{t}e^{-i(\delta _{R,3}+\delta _{R,4})t}$ for each
basis $|j\rangle _{t}$ can be applied to eliminate the time-dependency of
SOC, yielding the effective Hamiltonian
\begin{equation}
H_{5,0}=\sum_{j=1}^{5}\frac{\hat{p}^{2}}{2}|j\rangle \langle
j|+\sum_{j=1}^{4}\left( \frac{\Omega }{2}e^{-2i\hat{k}_{j}\cdot \hat{r}%
}|j\rangle \langle j+1|+h.c.\right) .
\end{equation}%
where the laser strengths are chosen such that each SOC term has the same
effective Raman coupling strength $\Omega =J\left\vert c_{p,j}\right\vert
\Omega _{j}/\Delta _{j}$. The detunings are canceled out for exactly
resonant case.

\subsection{$k$-space configuration}

\begin{figure}[t]
\centering
\includegraphics[width=0.45\textwidth]{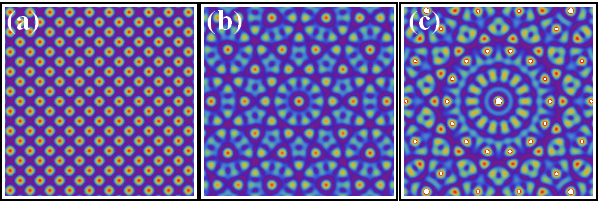}
\caption{(Color online) Crystalline and quasicrystal patterns in real-space
when the minima in $k$-space form a regular polygon and are evenly
populated. (a) $n=4$, namely, a square in momentum space. Discrete
translational symmetries in both $x$ and $y$ directions are preserved. (b)
and (c) correspond to $n=5$ (regular pentagon) and $n=7$ (heptagon)
respectively. Discrete translational symmetries are broken and corresponding
rotational symmetries are exhibited.}
\label{figS1}
\end{figure}

The spatial dependence of the Raman coupling can be removed by a unitary
transformation
\begin{equation}
|j\rangle \rightarrow e^{i\hat{l}_{j}\cdot \hat{r}}|j\rangle ,\text{ for }%
j=1...5,
\end{equation}%
for each state, where $\hat{l}_{j}$ is a constant vector. In this new
pseudo-momentum basis, $\hat{l}_{j}$ must satisfy the following group of
equations%
\begin{equation}
-\hat{l}_{j}-2\hat{k}_{j}+\hat{l}_{j+1}=0,\text{ for }j=1...4.
\end{equation}%
to eliminate the spatial dependence of Raman couplings. However, these four
equations are not sufficient to determine all variables. The fifth equation
can be obtained from minimizing the single particle energy functional.
Considering the simplest case $\Omega =0$ with the kinetic energy given by%
\begin{equation}
E_{k,0}=\frac{1}{2}\sum_{j=1}^{5}\hat{l}_{j}^{2}=\frac{1}{2}\hat{l}%
_{1}^{2}+\sum_{j=2}^{5}\left( \hat{l}_{1}+2\sum_{j-1}^{4}\hat{k}_{j}\right)
^{2}.
\end{equation}%
Minimizing this functional yields $\hat{l}_{1}=-\frac{2}{5}%
\sum_{j=1}^{4}(5-j)\hat{k}_{j}$. $\hat{k}_{j}$ is exactly the same as that
in the main text for a pentagon in momentum space with one point fixed at $%
(0,1)$. Inserting $\hat{k}_{j}$ back, one finds $\hat{l}_{1}=(0,1)$, which
is consistent with our configuration. Therefore $\hat{l}_{j}$ are nothing
but the minima in $k$-space, that is, $\hat{l}_{j}=\hat{p}_{j}$. This is not
true for arbitrary configuration but holds for any regular polygon. If all
minima are occupied equally, we may have crystal or quasicrystal orders. In
Fig.~\ref{figS1}, we illustrate the spatial distribution patterns for
typical crystal and quasicrystal orders. Panel (a) is a simple crystal with
four band minima and discrete lattice translational symmetry. (b)[(c)] is a
quasicrystal with 5-fold (7-fold) rotational symmetry if the phase
distribution is taken into account. Discrete lattice translational symmetry
is obviously absent. For any $n\geq 7$, we may expect a quasicrystal
structure in its spatial pattern.

\subsection{Non-zero-momentum plane-wave phase}

With interactions, the plane-wave phase would occur at a large Raman
coupling. The Hamiltonian can be rewritten as
\begin{equation}
H_{5,0}^{\prime }=\Omega \left( \frac{1}{\Omega }H_{p}+H_{soc}\right) ,
\end{equation}%
where $H_{p}=\sum_{j=1}^{n}\frac{1}{2}(p-p_{j})^{2}|j\rangle \langle j|$ is
the kinetic energy and $H_{soc}=\sum_{j=1}^{n-1}\left( |j\rangle \langle
j+1|+H.c.\right) $. In the large $\Omega $ region, the kinetic energy term
can be treated as a perturbation. The first-order energy correction is
\begin{equation}
E_{0}^{(1)}=\langle \Phi _{0}|\frac{1}{\Omega }H_{p}|\Phi _{0}\rangle =\frac{%
1}{2\Omega }\sum_{j}|a_{j}|^{2}(p-p_{j})^{2},
\end{equation}%
where $\Phi _{0}=\sum_{j}a_{j}|j\rangle $ is the ground state of $H_{soc}$
with the normalization $\sum_{j}|a_{j}|^{2}=1$. A simple variational method
shows that the minimum locates at $\sum_{j}|a_{j}|^{2}p_{j}$, which is
generally non-zero, yielding a non-zero-momentum plane-wave phase.


\begin{thebibliography}{99}
\bibitem{Christian1994} J. Christian, {Neutron and Synchrotron Radiation for
Condensed Matter Studies}, Springer Berlin Heidelberg, 1994.

\bibitem{Levine1984} D. Levine and P. J. Steinhardt, {Quasicrystals: A New
Class of Ordered Structures}, \href{https://doi.org/10.1103/PhysRevLett.53.2477}%
{Phy. Rev. Lett. \textbf{53}, 2477 (1984)}.

\bibitem{Shechtman1984} D. Shechtman, I. Blech, D. Gratias, and J. W. Cahn, {%
Metallic Phase with Long-Range Orientational Order and No Translational
Symmetry}, \href{https://doi.org/10.1103/PhysRevLett.53.1951}{Phy. Rev.
Lett. \textbf{53}, 1951 (1984)}.

\bibitem{Macia2006} E. Maci\'{a}, {The role of aperiodic order in science
and technology}, \href{https://doi.org/doi:10.1088/0034-4885/69/2/R03}{Rep.
Prog. Phys. \textbf{69}, 397 (2006)}.

\bibitem{Freedman2007} B. Freedman, R. Lifshitz, J. W. Fleischer \& M.
Segev, {Phason dynamics in nonlinear photonic quasicrystals}, \href{https://doi.org/doi:10.1038/nmat1981}%
{Nat. Mater. \textbf{6}, 776 (2007)}.

\bibitem{Barkan2011} K. Barkan, H. Diamant, and R. Lifshitz, {Stability of
quasicrystals composed of soft isotropic particles}, \href{https://doi.org/10.1103/PhysRevB.83.172201}%
{Phy. Rev. B \textbf{83}, 172201 (2011)}.

\bibitem{Wasio2014} N. A. Wasio, R. C. Quardokus, R. P. Forrest, C. S. Lent,
S. A. Corcelli, J. A. Christie, K. W. Henderson \& S. Alex Kandel, {%
Self-assembly of hydrogen-bonded two-dimensional quasicrystals}, \href{https://doi.org/doi:10.1038/nature12993}%
{Nature (London) \textbf{507}, 86 (2014)}.

\bibitem{Nagao2015} K. Nagao, T. Inuzuka, K. Nishimoto, and K. Edagawa, {%
Experimental Observation of Quasicrystal Growth}, \href{https://doi.org/10.1103/PhysRevLett.115.075501}%
{Phy. Rev. Lett. \textbf{115}, 075501 (2015)}.

\bibitem{Urgel2016} J. I. Urgel, D. \'{E}cija, G. Lyu, R. Zhang, C.-A.
Palma, W. Auw$\ddot{\text{a}}$rter, N. Lin \& J. V. Barth, {%
Quasicrystallinity expressed in two-dimensional coordination networks},
\href{https://doi.org/doi:10.1038/nchem.2507}{Nat. Chem. \textbf{8}, 657
(2016)}.

\bibitem{Bindi2009} L. Bindi, P. J. Steinhardt, N. Yao, P. J. Lu, {Natural
Quasicrystals}, \href{https://doi.org/10.1126/science.1170827}{Science
\textbf{324}, 1306 (2009)}.


\bibitem{Boninsegni2012} M. Boninsegni and N. V. Prokofv, {Supersolids: What
and where are they?} \href{https://doi.org/10.1103/RevModPhys.84.759}{Rev.
Mod. Phys. \textbf{84}, 759 (2012)}.

\bibitem{Thouless1969} D.J Thouless, {The flow of a dense superfluid}, \href{https://doi.org/10.1016/0003-4916(69)90286-3}%
{Ann. Phys. \textbf{52}, 403 (1971)}.

\bibitem{Andreev1971} A. F. Andreev and I. M. Lifshitz, {Quantum Theory of
Defects in Crystals}, \href{https://doi.org/10.1070/PU1971v013n05ABEH004235}{%
Sov. Phys. JETP \textbf{29}, 1107 (1971)}.

\bibitem{Li2017} J.-R. Li, J. Lee, W. Huang, S. Burchesky, B. Shteynas, F.
\c{C}. Top, A. O. Jamison, and W. Ketterle, {A stripe phase with supersolid
properties in spin-orbit-coupled Bose-Einstein condensates}, \href{https://doi.org/doi:10.1038/nature21431}%
{Nature (London) \textbf{543}, 91 (2017)}.

\bibitem{Leonard2017} J. L\'{e}onard, A. Morales, P. Zupancic, T. Esslinger
\& T. Donner , {Supersolid formation in a quantum gas breaking a continuous
translational symmetry}, \href{https://doi.org/doi:10.1038/nature21067}{%
Nature (London) \textbf{543}, 87 (2017)}.


\bibitem{Li2016} J. Li, W. Huang, B. Shteynas, S. Burchesky, F. Top, E. Su,
J. Lee, A. O. Jamison, and W. Ketterle, {Spin-Orbit Coupling and Spin
Textures in Optical Superlattices}, \href{https://doi.org/10.1103/PhysRevLett.117.185301}%
{Phy. Rev. Lett. \textbf{117}, 185301 (2016)}.


\bibitem{MyApp} Supplementary materials, see supplementary materials for
details.

\bibitem{Luo2017} X.-W. Luo, K. Sun, C. Zhang, Spin-tensor--momentum-coupled
Bose-Einstein condensates, Phys. Rev. Lett. \textbf{119}, 193001 (2017).


\bibitem{Abeelen1999} F. A. van Abeelen and B. J. Verhaar, {Determination of
collisional properties of cold Na atoms from analysis of bound-state
photoassociation and Feshbach resonance field data}, \href{https://doi.org/10.1103/PhysRevA.59.578}%
{Phy. Rev. A \textbf{59}, 578 (1999)}.

\bibitem{Knoop2011} S. Knoop, T. Schuster, R. Scelle, A. Trautmann, J.
Appmeier, M. K. Oberthaler, E. Tiesinga, and E. Tiemann, {Feshbach
spectroscopy and analysis of the interaction potentials of ultracold sodium}%
, \href{https://doi.org/10.1103/PhysRevA.83.042704}{Phy. Rev. A \textbf{83},
042704 (2011)}.

\bibitem{Lin2011} Y.-J. Lin, K. Jim\'{e}nez-Garc\'{\i}a, and I. B. Spielman,
{Spin-orbit-coupled Bose-Einstein condensates}, \href{http://dx.doi.org/10.1038/nature09887}%
{Nature (London) \textbf{471}, 83 (2011)}.

\bibitem{Zhang2012b} J.-Y. Zhang, S.-C. Ji, Z. Chen, L. Zhang, Z.-D. Du, B.
Yan, G.-S. Pan, B. Zhao, Y.-J. Deng, H. Zhai, S. Chen, and J.-W. Pan, {%
Collective Dipole Oscillations of a Spin-Orbit Coupled Bose-Einstein
Condensate}, \href{http://dx.doi.org/10.1103/PhysRevLett.109.115301}{Phys.
Rev. Lett. \textbf{109}, 115301 (2012)}.

\bibitem{Qu2013a} C. Qu, C. Hamner, M. Gong, C. Zhang, and P. Engels, {%
Observation of Zitterbewegung in a spin-orbit-coupled Bose-Einstein
condensate}, \href{http://dx.doi.org/10.1103/PhysRevA.88.021604}{Phys. Rev.
A \textbf{88}, 021604(R) (2013)}.

\bibitem{Olson2014} A. J. Olson, S.-J. Wang, R. J. Niffenegger, C.-H. Li, C.
H. Greene, and Y. P. Chen, {Tunable Landau-Zener transitions in a
spin-orbit-coupled Bose-Einstein condensate}, \href{http://dx.doi.org/10.1103/PhysRevA.90.013616}%
{Phys. Rev. A \textbf{90}, 013616 (2014)}.

\bibitem{Hamner2014} C. Hamner, C. Qu, Y. Zhang, J. Chang, M. Gong, C.
Zhang, and P. Engels, {Dicke-type phase transition in a spin-orbit-coupled
Bose-Einstein condensate}, \href{http://dx.doi.org/10.1038/ncomms5023}{Nat.
Commun. \textbf{5}, 4023 (2014)}.

\bibitem{Wang2012} P. Wang, Z.-Q. Yu, Z. Fu, J. Miao, L. Huang, S. Chai, H.
Zhai, and J. Zhang, {Spin-Orbit Coupled Degenerate Fermi Gases}, \href{http://dx.doi.org/10.1103/PhysRevLett.109.095301}%
{Phys. Rev. Lett. \textbf{109}, 095301 (2012)}.

\bibitem{Cheuk2012} L. W. Cheuk, A. T. Sommer, Z. Hadzibabic, T. Yefsah, W.
S. Bakr, and M. W. Zwierlein, {Spin-Injection Spectroscopy of a Spin-Orbit
Coupled Fermi Gas}, \href{http://dx.doi.org/10.1103/PhysRevLett.109.095302}{%
Phys. Rev. Lett. \textbf{109}, 095302 (2012)}.

\bibitem{Williams2013} R. A. Williams, M. C. Beeler, L. J. LeBlanc, K. Jim%
\'{e}nez-Garc\'{\i}a, and I. B. Spielman, {Raman-Induced Interactions in a
Single-Component Fermi Gas Near an $s$-Wave Feshbach Resonance}, \href{http://dx.doi.org/10.1103/PhysRevLett.111.095301}%
{Phys. Rev. Lett. \textbf{111}, 095301 (2013)}.

\bibitem{Lev} N. Q. Burdick, Y. Tang, and B. L. Lev, {Long-Lived
Spin-Orbit-Coupled Degenerate Dipolar Fermi Gas}, \href{https://doi.org/10.1103/PhysRevX.6.031022}%
{Phys. Rev. X \textbf{6}, 031022 (2016)}.

\bibitem{Jo} B. Song, C. He, S. Zhang, E. Hajiyev, W. Huang, X.-J. Liu, and
G.-B. Jo, {Spin-orbit-coupled two-electron Fermi gases of ytterbium atoms},
\href{https://doi.org/10.1103/PhysRevA.94.061604}{Phys. Rev. A \textbf{94},
061604(R) (2016)}.

\bibitem{Huang2016} L. Huang, Z. Meng, P. Wang, P. Peng, S.-L. Zhang, L.
Chen, D. Li, Q. Zhou \& J. Zhang, {Experimental realization of
two-dimensional synthetic spin--orbit coupling in ultracold Fermi gases},
\href{https://doi.org/doi:10.1038/nphys3672}{Nat. Phys. \textbf{12}, 540
(2016)}.

\bibitem{Meng2016} Z. Meng, L. Huang, P. Peng, D. Li, L. Chen, Y. Xu, C.
Zhang, P. Wang, and J. Zhang, {Experimental Observation of a Topological
Band Gap Opening in Ultracold Fermi Gases with Two-Dimensional Spin-Orbit
Coupling}, \href{https://doi.org/10.1103/PhysRevLett.117.235304}{Phys. Rev.
Lett. \textbf{117}, 235304 (2016)}.

\bibitem{Pan2016} Z. Wu, L. Zhang, W. Sun, X.-T. Xu, B.-Z. Wang, S.-C. Ji,
Y. Deng, S. Chen, X.-J. Liu, J.-W. Pan, {Realization of two-dimensional
spin-orbit coupling for Bose-Einstein condensates}, \href{https://doi.org/10.1126/Science.aaf6689}%
{ Science \textbf{354}, 83 (2016)}.

\bibitem{Stanescu2008} T.~D.~Stanescu, B.~Anderson, and V.~Galitski, %
\newblock {Spin-orbit coupled Bose-Einstein condensates}, \newblock \href{http://dx.doi.org/10.1103/PhysRevA.78.023616}%
{Phys. Rev. A \textbf{78}, 023616 (2008)}.

\bibitem{Wu2011} C.~Wu, I.~Mondragon-Shem, and X.-F.~Zhou, \newblock
{Unconventional Bose-Einstein Condensations from Spin-Orbit Coupling}, %
\newblock \href{http://dx.doi.org/10.1088/0256-307X/28/9/097102}{Chin. Phys.
Lett. \textbf{28}, 097102 (2011)}.

\bibitem{wang2010spin} C.~Wang, C.~Gao, C.-M. Jian, and H.~Zhai, \newblock %
Spin-orbit coupled spinor Bose-Einstein condensates, \newblock \href{http://dx.doi.org/10.1103/PhysRevLett.105.160403}%
{Phys. Rev. Lett. \textbf{105}, 160403 (2010)}.

\bibitem{ho2011bose} T.-L.~Ho and S.~Zhang, \newblock Bose-Einstein
condensates with spin-orbit interaction, \newblock \href{http://dx.doi.org/10.1103/PhysRevLett.107.150403}%
{Phys. Rev. Lett. \textbf{107}, 150403 (2011)}.

\bibitem{li2012quantum} Y.~Li, L.~Pitaevskii, and S.~Stringari, \newblock %
Quantum tricriticality and phase transitions in spin-orbit coupled
Bose-Einstein condensates, \newblock \href{http://dx.doi.org/10.1103/PhysRevLett.108.225301}%
{Phys. Rev. Lett. \textbf{108}, 225301 (2012)}.

\bibitem{zhang2012mean} Y.~Zhang, L.~Mao, and C.~Zhang, \newblock Mean-field
dynamics of spin-orbit coupled Bose-Einstein condensates, \newblock \href{http://dx.doi.org/10.1103/PhysRevLett.108.035302}%
{Phys. Rev. Lett. \textbf{108}, 035302 (2012)}.

\bibitem{hu2012spin} H.~Hu, B.~Ramachandhran, H.~Pu, and X.-J.~Liu, %
\newblock Spin-orbit coupled weakly interacting Bose-Einstein condensates in
harmonic traps, \newblock \href{http://dx.doi.org/10.1103/PhysRevLett.108.010402}%
{Phys. Rev. Lett. \textbf{108}, 010402 (2012)}.

\bibitem{ozawa2012stability} T.~Ozawa and G.~Baym, \newblock Stability of
ultracold atomic bose condensates with rashba spin-orbit coupling against
quantum and thermal fluctuations, \newblock \href{http://dx.doi.org/10.1103/PhysRevLett.109.025301}%
{Phys. Rev. Lett. \textbf{109}, 025301 (2012)}.

\bibitem{sun2016interacting} K.~Sun, C.~Qu, Y.~Xu, Y.~Zhang, and C.~Zhang, %
\newblock Interacting spin-orbit-coupled spin-1 Bose-Einstein condensates, %
\newblock \href{http://dx.doi.org/10.1103/PhysRevA.93.023615}{Phys. Rev. A
\textbf{93}, 023615 (2016)}.

\bibitem{yu2016phase} Z.-Q.~Yu, \newblock Phase transitions and elementary
excitations in spin-1 Bose gases with Raman-induced spin-orbit coupling, %
\newblock \href{http://dx.doi.org/10.1103/PhysRevA.93.033648}{Phys. Rev. A
\textbf{93}, 033648 (2016)}.

\bibitem{martone2016tricriticalities} G.~Martone, F.~Pepe, P.~Facchi,
S.~Pascazio, and S.~Stringari, \newblock Tricriticalities and quantum phases
in spin-orbit-coupled spin-1 bose gases, \newblock\href{http://dx.doi.org/10.1103/PhysRevLett.117.125301}%
{Phys. Rev. Lett. \textbf{117}, 125301 (2016)}.

\bibitem{Chin2015} L. W. Clark, L.-C. Ha, C.-Y. Xu, and C. Chin, {Quantum
Dynamics with Spatiotemporal Control of Interactions in a Stable
Bose-Einstein Condensate}, \href{https://doi.org/10.1103/PhysRevLett.115.155301}%
{Phys. Rev. Lett. \textbf{115}, 155301 (2015)}.
\end{thebibliography}
\end{document}